\documentclass[a4paper, 12pt]{article}
\usepackage[utf8]{inputenc}
\usepackage[T2A]{fontenc}
\usepackage[english]{babel}
\usepackage{amssymb}
\usepackage{amsmath}
\usepackage{textcomp}
\usepackage{indentfirst}
\usepackage{float}
\usepackage{caption}

\usepackage[sorting=none,backend=bibtex,citestyle=numeric]{biblatex}
\bibliography{literature.bib}

\DeclareFieldFormat{pages}{#1}

\usepackage{units}
\usepackage{amsfonts}
\usepackage{graphicx} 
\usepackage{amstext}
\usepackage{wasysym} 
\usepackage{comment}
\usepackage{color}

\usepackage{authblk} 

\usepackage{caption}
\DeclareCaptionLabelSeparator{pipe}{ $: $ }
\usepackage{epstopdf}
\usepackage{xcolor}
\usepackage{hyperref}
 
\definecolor{linkcolor}{HTML}{000000}
\definecolor{urlcolor}{HTML}{000000}
 
\hypersetup{pdfstartview=FitH,  citecolor=black, linkcolor=linkcolor,urlcolor=urlcolor, colorlinks=true}
\usepackage[left=2.5cm,right=2cm, top=2cm,bottom=2cm,bindingoffset=0cm]{geometry}

\makeatletter
\newcommand{\l@abcd}[2]{\hbox to\textwidth{#1\dotfill #2}}
\renewcommand{\l@section}{\@dottedtocline{1}{2.5em}{2.3em}}
\renewcommand{\l@part}{\@dottedtocline{1}{1.5em}{2.3em}}
\renewcommand{\l@subsection}{\@dottedtocline{1}{3.5em}{2.3em}}
\makeatother

\title{Sound propagation and Mach cone in anisotropic hydrodynamics}

\author[1]{Martin Kirakosyan}
\author[1,2]{Aleksandr Kovalenko}
\author[1]{Andrey Leonidov}
\affil[1]{P.N. Lebedev Physical Institute, Moscow Russia}
\affil[2]{M.V. Lomonosov Moscow State University, Moscow, Russia}

\begin{document}

\maketitle

\begin{abstract}
This letter is based on a kinetic theory approach to anisotropic hydrodynamics. We derive the sound wave equation in anisotropic hydrodynamics and show that a corresponding wave front is ellipsoidal. The phenomenon of Mach cone emission in anisotropic hydrodynamics is studied. It is shown that Mach cone in anisotropic case becomes asymmetric, i. e. in this limit they're two different angles, left and right with respect to the ultrasonic particle direction, which are determined by the direction of ultrasonic particle propagation and the asymmetry coefficient.
\end{abstract}

\section{Introduction}

One of the universal features of highly excited matter created at the early stages of heavy ion collisions is its momentum space anisotropy, extreme at its birth point and, presumably, gradually disappearing in the course of its expansion. Of significant interest are therefore physical phenomena that are directly related to this anisotropy. A natural stylized framework for discussing such phenomena is the so-called anisotropic relativistic hydrodynamics  \cite{DenFlRybStr,MartStr} in which the momentum anisotropy of evolving "liquid"\; is built in explicitly. One of the most important phenomena in hydrodynamics is its sound excitation modes and, in particular, the related phenomenon of Mach cone. In this letter we analyze sound propagation and Mach cone emission in anisotropic relativistic hydrodynamics.

The interest to the Mach cone emission in the context of ultra relativistic heavy ion collisions (\cite{Shuryak, RoyChadhur,RenkRuppert}) was sparked, in particular, by the results on two particle correlations at RHIC (\cite{STAR,PHENIX}). An alternative explanation of the two-humped structure observed was in terms of Cherenkov radiation of a parton moving with velocity exceeding the speed of gluon propagation in the hot dense medium formed in heavy ion collisions (\cite{Dremin,DKLV}). However, subsequent studies at LHC did not confirm the existence of the visible two-bump structure in a fragmentation of away side jets (\cite{LHCATLAS,LHCCMS}) and, in addition, arguments for significant effects from  background suppression at RHIC results were spelled out (see, e.g., a detailed review in \cite{NMNSB}). It should also be noted that a two-hump structure may be explained by taking into account geometric fluctuations of initial state in heavy ion collisions (\cite{Jeon,MaWang}). Despite the fact that at present there is no direct experimental evidence for the Mach cone emission phenomenon, there is still a significant theoretical interest in adjusting a description of this universal phenomenon to the realistic stylized properties of the matter created at the early stages of heavy ion collisions, in particular, its momentum anisotropy. Moreover, there are other observables that could be affected by the Mach cone emission, i. e. the enhancement of low-pt particles away from the quenched jets (\cite{CMS2011,CMS2012, TacHir}).    

The analysis in this letter is based on a kinetic theory approach to anisotropic hydrodynamics  \cite{MartStr}. In the Section 2  we derive the sound wave equation in anisotropic hydrodynamics and show that a corresponding wave front becomes ellipsoidal. In the Section 3 we analyze the phenomenon of Mach cone emission in anisotropic hydrodynamics and show that the cone becomes asymmetric, i. e. there are two different angles, left and right with respect to the ultrasonic particle direction, which are determined by the direction of ultrasonic particle propagation and the asymmetry coefficient.

\section{Sound in anisotropic hydrodynamics}

The phenomenon of sound propagation is analyzed by studying the excitation modes in the linear approximation in fluctuations. Let us perform this analysis for the formulation of anisotropic hydrodynamics based on kinetic theory approach and use the Romatschke-Struckland ansatz (\cite{StrRom1,StrRom2,StrLect}) for massless gas:

\begin{equation}
f(x,p) = f_{iso}\Bigg( \frac{\sqrt{p^\mu \Xi_{\mu\nu}p^\nu}}{\Lambda(x)}\Bigg),
\label{Kin1}
\end{equation}
where  $ (p^\mu \Xi_{\mu\nu}p^\nu = \mathbf{p}^2 +\xi(x) p_\parallel^2)$  in the Landau rest frame LRF and  $\xi$ is an anisotropy parameter which is, generally speaking,  a function of the coordinates $x$.

From this simple model one can derive analytical expressions for particle number density \(n\) and components of energy-momentum tensor \(T^{\mu\nu}\) in the LRF using their standard definitions as the first and second moments of the distribution function \cite{StrRom1,StrRom2,StrLect}
\begin{align}
T^{00} &= \varepsilon (\Lambda,\xi) = R(\xi) \varepsilon_{iso} (\Lambda),\\
T^{11} &= T^{22} = P_\perp (\Lambda,\xi) = R_\perp (\xi) P_{iso} (\Lambda), \\
T^{33} &= P_\parallel (\Lambda,\xi) = R_\parallel (\xi) P_{iso} (\Lambda),\\
n(\Lambda,\xi)  &= \frac{n_{iso}(\Lambda)}{\sqrt{1+\xi}},
\label{f2}
\end{align}
where \(\varepsilon_{iso} (\Lambda), \ P_{iso} (\Lambda)\) and \(n_{iso}(\Lambda)\) are isotropic energy density, pressure and particle number density respectively obtained from isotropic distribution function \(f_{iso}\) and the dependence on the anisotropy parameter is factored out for all the variables considered. The dependence of the factors $R_{\; ,\perp,\parallel}$ on the  anisotropy parameter $\xi$ is given by (\cite{StrLect, FlRyb}): 
\begin{align}
R(\xi) &= \frac{1}{2} \Bigg( \frac{1}{1+\xi} + \frac{\arctan \sqrt{\xi}}{\sqrt{\xi}}\Bigg),\\
R_\perp(\xi) &= \frac{3}{2\xi} \Bigg( \frac{1 + (\xi^2-1)R(\xi)}{1+\xi}\Bigg),\\
R_\parallel(\xi) &= \frac{3}{\xi} \Bigg( \frac{(\xi+1)R(\xi) -1}{1+\xi}\Bigg).
\label{f3}
\end{align}

In what follows it turns out convenient to introduce the following parametrization for the quantities involved in describing the properties of an anisotropic fluid in terms of a four-vector $u_\mu(x)$ and a rapidity $\vartheta(x)$: 
\begin{align}
U^\mu &= (u_0 \cosh \vartheta, u_x, u_y, u_0 \sinh \vartheta),\\
X^\mu &= (u_\perp \cosh \vartheta, \frac{u_0 u_x}{u_\perp}, \frac{u_0 u_y}{u_\perp}, u_\perp \sinh \vartheta),\\
Y^\mu &= (0,-\frac{u_y}{u_\perp},\frac{u_x}{u_\perp},0),\\
Z^\mu &= (\sinh \vartheta, 0,0, \cosh \vartheta),
\label{basis}
\end{align}
where \(U^\mu\) is the 4-velocity vector that describes the hydrodynamic flow, \(Z^\mu\) defines the direction of the longitudinal axis and  \(X^\mu\) and \(Y^\mu\) define axes in the  transverse plane and the four-vector satisfies $u_\mu(x)$  \(u_0^2 = 1 + u_x^2 + u_y^2\) (4-vectors $U$ are normalized so that $U_{\mu}U^{\mu} = 1$, the same holds for $X, Y$ and $Z$), \(\vartheta\) is the fluid rapidity.  In the LRF one has $U^{\mu} = (1,0,0,0)$, $Z^{\mu} = (0,0,0,1)$

Assuming conservation of energy-momentum tensor \(T^{\mu\nu}\) and particle current \(j^\mu\) one gets:
\begin{align}
\partial_\mu T^{\mu\nu} = 0 \label{f41}\\
\partial_\mu j^\mu = 0.
\label{f42}
\end{align}
where 
\begin{align}
T^{\mu\nu} &= (\varepsilon + P_\perp)U^\mu U^\nu - P_\perp g^{\mu\nu} - (P_\perp - P_\parallel)Z^\mu Z^\nu,
\label{T1}\\
j^\mu &= nU^\mu.
\label{T2}
\end{align}

Linearization of (\ref{f41},\ref{f42}) leads to equations describing propagation of sound (\cite{landau1987fluid}). To derive sound equation for the case under consideration let us expand the 4-velocities $U$, $Z$, particle number density, energy and momentum densities with respect to a temperature parameter \(\Lambda(x)=\Lambda^{(0)} + \Lambda^{(1)}(x)\) to the leading order in gradients:

\begin{align}
U^\mu &= U^{(0)\mu} + U^{(1)\mu}(x),\\
Z^\mu &= Z^{(0)\mu} + Z^{(1)\mu}(x),\\
\varepsilon &= \varepsilon^{(0)} + \varepsilon^{(1)}(x),\\
P_{\perp,\parallel} &= P_{\perp,\parallel}^{(0)} + P_{\perp,\parallel}^{(1)}(x),\\
n &= n^{(0)} + n^{(1)}(x).
\label{linear}
\end{align}

Let us now move to the frame in which  \(U^{(0)} = (1,0,0,0)\) and \(Z^{(0)} = (0,0,0,1)\). That gives us, through \(U^\mu Z_\mu = 0\), same relations between components of \(U^{(1)}\) and \(Z^{(1)}\). Let us also introduce the following notations:
\begin{align}
\varepsilon^{(1)}(x) & = c_{\varepsilon} n^{(1)} (x),\\ 
P_{\perp}^{(1)}(x) & = c_{\perp} n^{(1)} (x),\\
P_{\parallel}^{(1)} & = c_{\parallel} n^{(1)} (x)
\end{align}

Let us also assume the smallness of the gradients of the anisotropy parameter \(\xi \). Limiting our consideration to transverse fluctuations with respect to $u_{\mu}^{(0)}$  we obtain, using \eqref{Kin1}, \eqref{f41} and \eqref{f42}, the following sound equation
\begin{equation}
\big( c_\perp \partial^2_x + c_\perp \partial^2_y + c_\parallel \partial^2_z - c_\varepsilon \partial^2_t \big)n^{(1)} = 0,
\label{wave1}
\end{equation}
where
\begin{align}
c_\varepsilon &= \sqrt{1+\xi}R(\xi)A \nonumber \\
c_\perp &= \sqrt{1+\xi}R_\perp(\xi)\frac{A}{3}, \nonumber \\
c_\parallel &= \sqrt{1+\xi}R_\parallel(\xi)\frac{A}{3},
\label{coefficients}
\end{align}

Introducing a notation \(\kappa = c_\parallel/c_\perp\) and using a relation \(c_\varepsilon = 2c_\perp + c_\parallel\) valid in the ultra-relativistic case , we can rewrite \eqref{wave1} in the following form:

\begin{equation}
\big(\partial^2_x + \partial^2_y + \kappa \partial^2_z \big)n^{(1)} = (\kappa + 2) \partial^2_t n^{(1)}.
\label{wave2}
\end{equation}
In the isotropic limit \(\kappa =1 \)  equation \eqref{wave2} reduces to the standard isotropic ultrarelativistic sound equation.  Let us note that  \(\kappa\) may be written as \(\kappa = \frac{P_\parallel}{P_\perp}\) and, therefore, \(\kappa\) is the anisotropy-related quantity that is, in principle, observable. A dependence of   \(\kappa\) on the anisotropy parameter is shown in Fig.~ 1.

\begin{figure}[H]
	\center{\includegraphics[width=0.6\linewidth]{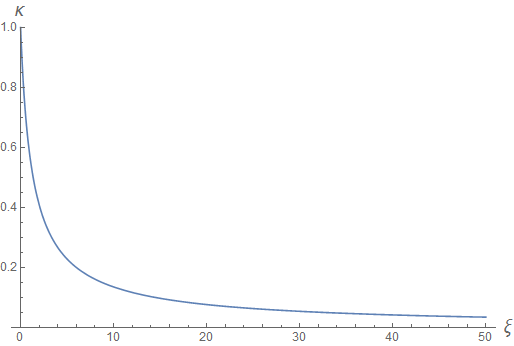}}
	\caption{\(\kappa(\xi)\).}
	\label{ris1}
\end{figure}

\section{Mach cone}

One of the characteristic phenomena in hydrodynamics is the appearance of the Mach cone, an expanding shock wave that is generated by an ultrasonic body propagating in a medium. Its properties depend on the ratio of the local flow velocity \(v\) and the speed of sound in the medium \(c_s\) - the Mach Number (MN). The Mach cone appears for $MN>1$.  For isotropic theories there is a well known formula for Mach angle, an angle with respect to the direction of propagation of ultrasonic particle at which the shock wave is emitted: \(\sin\theta_M = c_s/v\). Obviously, in anisotropic theory there is no such simple relation. First,  the base of the cone is no longer a circle, but an ellipse, i.e. the shape of the sound wave front is ellipsoidal.  Second, the front is symmetric in a plane transversal to direction of anisotropy, in the wave equation (\ref{wave2}) it is \(Oxy\) plane. Thus, one may consider 2D-problem instead of 3D-problem and fix any axis in the transverse plane (say \(Ox\)). In Figure 2 we plot a 2-D slice of a full 3-D picture and show a particle moving with velocity \(v\) from \(O\) in the plane \(Oxz\) at an angle \(\alpha\) to the \(Ox\) axis. With  \(v/c_s > 1\)  there appears a Mach cone, which is formed by tangents to the ellipsoid. In Figure 2 the particle is at point \(B\) and \(AB\), \ \(CB\) are tangents to the ellipsoid, \(\theta_{MR}, \ \theta_{ML}\) are the Mach angles.

\begin{figure}[H]
	\center{\includegraphics[width=0.6\linewidth]{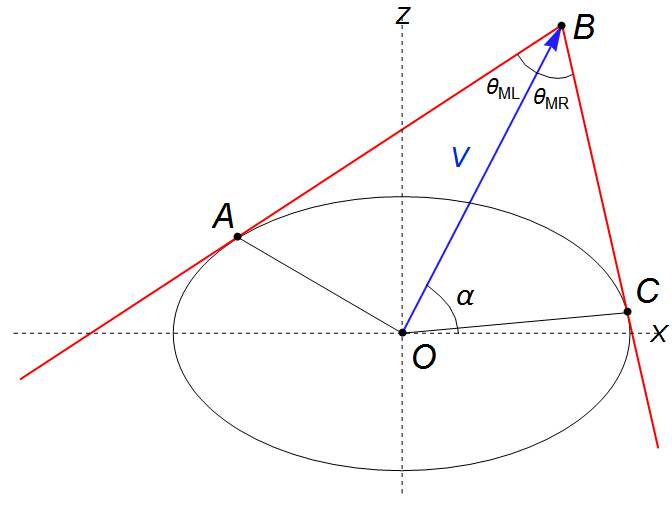}}
	\caption{Representation of the Mach cone}
	\label{ris1}
\end{figure}

Having an anisotropic sound with \(c_z = \sqrt{\kappa}/\sqrt{\kappa+2}, \ c_x = 1/\sqrt{\kappa+2}\) we get the following formula for the ellipse (say, upper part \(z>0\) ) and tangent \(AB\)  as functions of \(x\):
\begin{align}
f_1(x) &= c_z \sqrt{1 - \frac{x^2}{c_x^2}},\\
f_2(x) &= v \sin\alpha + (x - v\cos\alpha)\tan\Big(\alpha  - \theta_M\Big).
\end{align}
Equating them one gets a  quadratic equation for \(x\), and an equation describing the Mach cone follows from the condition of existence of its roots. We get
\begin{align}
\theta_{ML} &= \arctan\Bigg[\frac{\sin \alpha\cos\alpha(\kappa - 1) + \sqrt{(\kappa\cos^2\alpha + \sin^2\alpha)(\kappa+2) - \kappa}}{(\kappa+2) - \kappa\sin^2\alpha - \cos^2\alpha}\Bigg],
\\
\theta_{MR} &= \arctan\Bigg[\frac{\sin \alpha\cos\alpha(1-\kappa) + \sqrt{(\kappa\cos^2\alpha + \sin^2\alpha)(\kappa+2) - \kappa}}{(\kappa+2) - \kappa\sin^2\alpha - \cos^2\alpha}\Bigg].
\end{align}

Here we define two different Mach angels \(\theta _{ML}, \ \theta _{MR}\), which characterize the whole Mach cone (not only its 2D-slice). It should be noted that if \(\kappa = 1\) (no anisotropy) then \(\theta _{ML} = \theta _{MR} = \arctan 1/\sqrt{2}\), the standard expression for the Mach angle in the relativistic isotropic hydrodynamics.

\begin{figure}[H]
	\center{\includegraphics[width=0.6\linewidth]{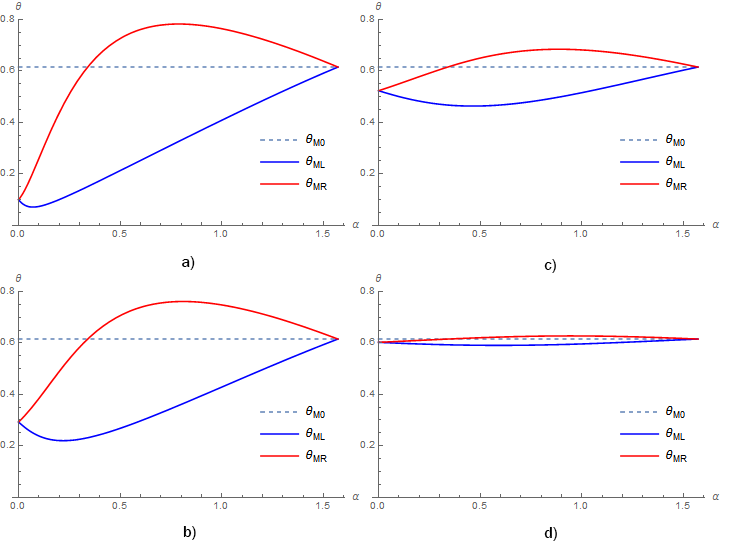}}
	\caption{Mach angles \(\theta_{ML}\) and \(\theta_{MR}\) as functions of \(\alpha\), \(\theta_{M0}\) is a Mach angle in isotropic case. \(a)\): \(\kappa =0.01\), \(b)\): \(\kappa =0.1\), \(c)\): \(\kappa =0.5\), \(d)\): \(\kappa =0.9\)}
	\label{ris1}
\end{figure}
\begin{figure}[H]
	
	\center{\includegraphics[width=0.6\linewidth]{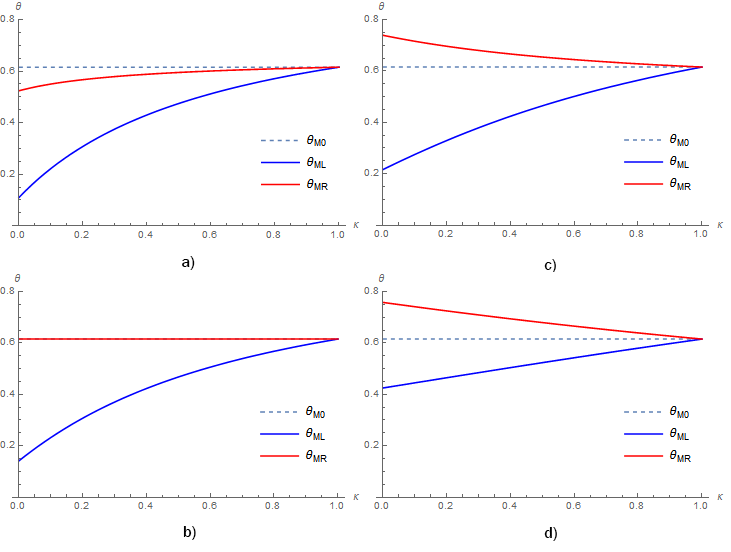}}
	\caption{Mach angles \(\theta_{ML}\) and и \(\theta_{MR}\) as functions of \(\kappa\), \(\theta_{M0}\) is a Mach angle in isotropic case. \(a)\): \(\alpha =\pi/12\), \(b)\): \(\alpha = \arcsin 1/3\), \(c)\): \(\alpha = \pi/6\), \(d)\): \(\alpha=\pi/3\)}
	\label{ris1}
\end{figure}

\section{Conclusions}

In this letter we have developed an analytical description of the Mach cone in relativistic anisotropic hydrodynamics.

\printbibliography

\end{document}